\documentclass[conference,10pt]{IEEEtran}
\usepackage{amsmath,graphicx}
\usepackage{pgf,tikz,pgfplots}
\usepackage{cite}
\usepackage{amsmath,amssymb,amsthm}
\usepackage{algorithmic}
\usepackage[linesnumbered,boxed]{algorithm2e}
\usepackage{bm}
\usepackage{color}
\newtheorem{theorem}{Theorem}

\usepackage{subfigure}

\newcommand{\liuhao}{\fontsize{8pt}{\baselineskip}\selectfont}

\def\notvartualgraph{1}
\ifthenelse{\notvartualgraph=1}{

}
{

}

\begin{document}
\title{Accelerating Coordinate Descent via Active Set Selection for Device Activity Detection for Multi-Cell Massive Random Access}
\author{\IEEEauthorblockN{Ziyue Wang\IEEEauthorrefmark{1}\IEEEauthorrefmark{2},
		Ya-Feng Liu\IEEEauthorrefmark{2},
		Zhilin Chen\IEEEauthorrefmark{3},
		{Wei Yu\IEEEauthorrefmark{3}}
	}
	\IEEEauthorblockA{\IEEEauthorrefmark{1}School of Mathematical Sciences, University of Chinese Academy of Sciences, Beijing, China}
	\IEEEauthorblockA{\IEEEauthorrefmark{2}LSEC, ICMSEC, Academy of Mathematics and Systems Science, Chinese Academy of Sciences, Beijing, China}
	\IEEEauthorblockA{\IEEEauthorrefmark{3}Department of Electrical and Computer Engineering, University of Toronto, Toronto, Canada}
	\small{Email: wangziyue20@mails.ucas.ac.cn, yafliu@lsec.cc.ac.cn, \{zchen, weiyu\}@ece.utoronto.ca}}

\setlength{\abovedisplayskip}{0.08cm}
\setlength{\belowdisplayskip}{0.08cm}
\setlength{\jot}{0.08cm}

\maketitle


\begin{abstract}
	We propose a computationally efficient algorithm for the device activity detection problem in the multi-cell massive multi-input multi-output (MIMO) system, where the active devices transmit their signature sequences to multiple BSs in multiple cells and all the BSs cooperate to detect the active devices. The device activity detection problem has been formulated as a maximum likelihood maximization (MLE) problem in the literature. The state-of-the-art algorithm for solving the problem is the (random) coordinate descent (CD) algorithm. However, the CD algorithm fails to exploit the special sparsity structure of the solution of the device activity detection problem, i.e., most of devices are not active in each time slot. In this paper, we propose a novel active set selection strategy to accelerate the CD algorithm and propose an efficient active set CD algorithm for solving the considered problem. Specifically, at each iteration, the proposed active set CD algorithm first selects a small subset of all devices, namely the \emph{active set}, which contains a few devices that contribute the most to the deviation from the first-order optimality condition of the MLE problem thus potentially can provide the most improvement to the objective function, then applies the CD algorithm to perform the detection for the devices in the active set. Simulation results show that the proposed active set CD algorithm significantly outperforms the state-of-the-art CD algorithm in terms of the computational efficiency.
\end{abstract}


\section{Introduction}

Massive machine-type communication (mMTC) is expected to play a crucial role in the fifth-generation (5G) and beyond cellular systems \cite{Bockelmann2016}. A challenge in mMTC is massive random access, in which a large number of devices with sporadic data traffic wish to connect to the network in the uplink \cite{ChenX2021}. This task can be accomplished by a pilot stage, during which the active devices transmit their preassigned non-orthogonal signature sequences, and the network then identifies the active devices based on the received signals at the base-stations (BSs) \cite{Liu2018b}. The non-orthogonality of the sequences inevitably causes both intra-cell and inter-cell interference that impair the detection performance. This paper considers a cloud radio-access network (C-RAN) architecture to alleviate the inter-cell interference, and proposes a novel device activity detection algorithm, which is more computationally efficient than the state-of-the-art algorithm.

The device activity detection problem can be formulated as a compressed sensing problem, in which the instantaneous channel state information (CSI) and the device activity are jointly recovered by exploiting the sparsity in the device activity \cite{Senel2018,Liu2018,Chen2018,liu2021efficient}. When the CSI is not needed and the BSs are equipped with a large number of antennas, it is also possible to jointly estimate the device activities and the channel large-scale fading components by exploiting the channel statistical information via maximum likelihood estimation (MLE). This approach is proposed in \cite{Haghighatshoar2018} and termed as the covariance approach because the detection relies on the sample covariance matrix of the received signal. As compared to the compressed sensing approach, this covariance approach has the advantage of detecting many more active devices due to its quadratic-like scaling law \cite{Fengler2019a}. While \cite{Haghighatshoar2018} considers the covariance approach in the single-cell scenario, the extensions to the multi-cell systems are studied in \cite{Ganesan2020, chen2021sparse}, where the large-scale fading components are further assumed available and the cooperation among BSs are allowed.

In the covariance approach, the coordinate descent (CD) algorithm that iteratively updates the activity indicator of each device is commonly used since it can achieve excellent detection performance; see \cite{Haghighatshoar2018,ChenZ2020,Jiang2020a} for more details. In the single-cell scenario, the CD algorithm is also computationally efficient because it admits a closed-form solution for each update. Moreover, the computational efficiency of CD can further be improved by carefully designing the sampling strategy for coordinate selection \cite{Dong2020} or employing the active set technique to update only a subset of the coordinates \cite{wang2021efficient}. However, in the multi-cell scenario with BS cooperation, the CD algorithm becomes less appealing as it is generally impossible to get a closed-form solution for each coordinate update, which involves a polynomial function whose degree depends on the number of BSs \cite{Ganesan2020, chen2021sparse}.

In this paper, we propose a computationally efficient active set algorithm for solving the activity detection problem in the mulit-cell scenario. By noting that most of devices are inactive, we exploit the sparsity in the device activity to reduce the computations on the inactive devices to improve the computational efficiency.
To this end, with the same motivation as in \cite{wang2021efficient}, we design a novel active set selection strategy to accelerate the CD algorithm. Specifically, at each iteration, the proposed active set CD algorithm first selects a small subset of all devices, termed as the \emph{active set}, which contains a few devices that contribute the most to the deviation from the first-order optimality condition of the optimization problem thus potentially can provide the most improvement to the overall objective function, then applies the CD algorithm to perform the detection for the devices in the active set. Simulation results show that the proposed active set CD algorithm significantly outperforms the state-of-the-art CD algorithm in terms of the computational efficiency without any detection performance loss.
We also establish the finite termination property of the proposed active set CD algorithm.


%
%
\section{System Model and Problem Formulation}
\label{sec:modelformulation}
%
%
%
%
\subsection{System Model}
Consider an uplink massive MIMO multi-cell system consisting of $B$ cells. Each cell contains one BS equipped with $M$ antennas and $N$ devices each equipped with a single antenna. We assume a C-RAN architecture, in which all $B$ BSs are connected to a central unit (CU) via fronthaul links such that the received signals can be collected and jointly processed at the CU for inter-cell interference mitigation. Let $\sqrt{g_{ibn}}\mathbf{h}_{ibn}$ denote the channel between device $n$ in cell $b$ and BS $i,$ where $g_{ibn} \geq 0$ is the large-scale fading coefficient depending on path-loss and shadowing, and $\mathbf{h}_{ibn}\in\mathbb{C}^{M}$ is the Rayleigh fading component following $\mathcal{CN}(\mathbf{0},\mathbf{I})$. Assume that only $K\ll N$ devices are active during any coherence interval. Let $a_{bn}$ indicate the activity of device $n$ in cell $b,$ i.e., $a_{bn}=1$ if the device is active and $a_{bn}=0$ otherwise. 
For device identification, each device is preassigned a unique signature sequence $\mathbf{s}_{bn}\in\mathbb{C}^{L}$ with $L$ being the sequence length.

In the uplink pilot stage, all active devices transmit their signature sequences as random access requests. Assuming that the sequences are transmitted synchronously, the received signal at BS $b$ can be expressed as
\begin{align}\label{eq.sys}
	\mathbf{Y}_b&=\sum_{n=1}^{N}a_{bn}\mathbf{s}_{bn}g_{bbn}^{\frac{1}{2}}\mathbf{h}_{bbn}^T  + \sum_{j\neq b} \sum_{n=1}^N a_{jn}\mathbf{s}_{jn}g_{bjn}^{\frac{1}{2}}\mathbf{h}_{bjn}^T+\mathbf{W}_b\nonumber\\
	&=\mathbf{S}_b\mathbf{A}_b\mathbf{G}^{\frac{1}{2}}_{bb}\mathbf{H}_{bb}+\sum_{j\neq b}\mathbf{S}_j\mathbf{A}_j\mathbf{G}^{\frac{1}{2}}_{bj}\mathbf{H}_{bj}+\mathbf{W}_b.
\end{align}
In the above,  $\mathbf{S}_j=[\mathbf{s}_{j1},\ldots,\mathbf{s}_{jN}]\in\mathbb{C}^{L\times N}$ is the signature sequence matrix of the devices in cell $j,$ $\mathbf{A}_{j}=\operatorname{diag}\{a_{j1},\ldots,a_{jN}\}$ is a diagonal matrix that indicates the activity of the devices in cell $j,$ $\mathbf{G}_{bj}=\operatorname{diag}\{g_{bj1},\ldots,g_{bjN}\}$ contains the large-scale fading components between the devices in cell $j$ and BS $b,$ $\mathbf{H}_{bj}=[\mathbf{h}_{bj1},\ldots,\mathbf{h}_{bjN}]^T\in\mathbb{C}^{N\times M}$ is the Rayleigh fading channel between the devices in cell $j$ and BS $b,$ and $\mathbf{W}_b$ is the additive Gaussian noise that follows $\mathcal{CN}(\mathbf{0},\sigma_w^2\mathbf{I}),$ where $\sigma_w^2$ is the variance of the background noise normalized by the transmit power for notational simplicity.

Let $\underline{\mathbf{a}}=[\mathbf{a}_{1}^T,\ldots,\mathbf{a}_{B}^T]^T\in\mathbb{R}^{BN}$ indicate the activity of all the devices, where $\mathbf{a}_{j} = [a_{j1}, \ldots, a_{jN} ]^T$ denotes the diagonal entries of $\textbf{A}_j.$

\subsection{Problem Formulation}

In this paper, we consider the activity detection problem with known large-scale fading coefficients, i.e., we assume $\mathbf{G}_{bj},$ for all $b$ and $j$ are known at the BS. In this case, the activity detection problem is to detect the active devices in the system based on the received signals
$\mathbf{Y}_b,~b=1,\ldots,B,$ and is equivalent to estimating the activity indicator vector $\underline{\mathbf{a}}.$

As shown in \cite{chen2021sparse}, the above activity dectction problem can be mathematically formulated as the MLE problem.
Notice from \eqref{eq.sys} that for each BS $b,$ the received signals at the $M$ antennas are i.i.d.\ due to the fact that the Rayleigh
fading components are i.i.d.\ over the antennas. Let $\mathbf{y}_{bm}$ denote the received signal at the $m$-th antenna. Then one can show that $\mathbf{y}_{bm}\sim\mathcal{CN}(\mathbf{0},\boldsymbol\Sigma_b),$ where $\boldsymbol\Sigma_b$ is given by
\begin{align}\label{eq.sys.multi.cell}
	{\boldsymbol\Sigma_b = \frac{1}{M}\mathbb{E}\left[\mathbf{Y}_b\mathbf{Y}_b^H\right]=\sum_{j=1}^B\mathbf{S}_j\mathbf{A}_j\mathbf{G}_{bj}\mathbf{S}_j^H+\sigma_w^2\mathbf{I},}
\end{align}
and the likelihood function of $\mathbf{Y}_{b}$ can be expressed as
\begin{align}\label{eq.likelihood.multi}
	p(\mathbf{Y}_b \mid \underline{\textbf{a}})=\frac{1}{|\pi\boldsymbol\Sigma_b|^M}\exp{\left(-\operatorname{tr}\left(M\boldsymbol\Sigma_b^{-1}\hat{\boldsymbol\Sigma}_b\right)\right)},
\end{align}
where $\hat {\boldsymbol\Sigma}_b = \mathbf{Y}_b\mathbf{Y}_b^H/M$ is the sample covariance matrix computed by averaging over different antennas.
Since the received signals $\mathbf{Y}_{b},$ $b = 1,\ldots,B,$ are independent given $\underline{\mathbf{a}},$
the likelihood function of
$\mathbf{Y}_{1}, \ldots, \mathbf{Y}_{B}$ can then be written as
\begin{align*}
	p(\mathbf{Y}_1,\ldots,\mathbf{Y}_B\mid\underline{\textbf{a}})= \prod_{b=1}^B p(\mathbf{Y}_b\mid\underline{\textbf{a}}).
\end{align*}
The maximization of the log-likelihood function, which is equivalent to the minimization of $- \tfrac{1}{M} \log p(\mathbf{Y}_1,\ldots,\mathbf{Y}_B|~\underline{\textbf{a}}),$ can be formulated as
\begin{subequations}\label{eq.prob1.multi}
	\begin{alignat}{2}\label{eq.prob1.2.multi}
		&\underset{\underline{\textbf{a}}}{\operatorname{minimize}}    &\quad& \sum_{b=1}^B\left(\log\left|\boldsymbol\Sigma_b\right|+ \operatorname{tr}\left(\boldsymbol\Sigma_b^{-1}\hat{\boldsymbol\Sigma}_b\right)\right)\\
		&\operatorname{subject\,to} &      &a_{bn} \in [0,1], \,\forall \, b,n,
		\label{eq.prob1.3.multi}
	\end{alignat}
\end{subequations}
where the box constraint \eqref{eq.prob1.3.multi} is a continuous relaxation of the binary constraint $a_{bn} \in \{0,1\}.$ The approach of estimating activity devices based on solving problem \eqref{eq.prob1.multi} is called the covariance based approach, because the solution of problem \eqref{eq.prob1.multi} depends on the received signal $\mathbf{Y}_b$ only through its sample covariance $\hat {\boldsymbol\Sigma}_b.$ The above problem \eqref{eq.prob1.multi} looks similar to its single-cell counterpart at the first glance, but in reality problem \eqref{eq.prob1.multi} is much more difficult to solve.

Let $f(\underline{\textbf{a}})$ denote the objective function of problem \eqref{eq.prob1.multi}. Then, for any $b=1,2,\ldots,B,\, n=1,2,\ldots,N,$ the gradient of $f(\underline{\textbf{a}})$ with respect to $a_{bn}$ is
\begin{equation*} \label{gradienteq}
	\left[\nabla f(\underline{\textbf{a}})\right]_{bn} = \sum_{j=1}^B g_{jbn}
	\left( \mathbf{s}_{bn}^H\boldsymbol\Sigma_j^{-1}\mathbf{s}_{bn} - \mathbf{s}_{bn}^H\boldsymbol\Sigma_j^{-1}\hat{\boldsymbol\Sigma}_j\boldsymbol\Sigma_j^{-1}\mathbf{s}_{bn}
	\right).
\end{equation*}
The first-order optimality condition of problem \eqref{eq.prob1.multi} is
\begin{align}\label{eq.kkt}
	\left[\nabla f(\underline{\textbf{a}})\right]_{bn}
	\begin{cases}
		\geq 0, & \operatorname{if}~a_{bn} = 0;   \\
		\leq 0, & \operatorname{if}~a_{bn} = 1; \\
		= 0, &  \operatorname{if}~ 0 < a_{bn} < 1,
	\end{cases}\ \forall~b, n,
\end{align}
which is equivalent to
$$\operatorname{Proj}(\underline{\textbf{a}} - \nabla f(\underline{\textbf{a}})) - \underline{\textbf{a}} = \textbf{0},$$
where $\operatorname{Proj}(\cdot)$ is the projection operator onto the interval $[0,1].$

\section{Proposed Active Set CD Algorithm}

The basic idea of the proposed active set CD algorithm for solving problem \eqref{eq.prob1.multi} is to fully exploit the special structure of its solution, i.e., most components of the solution will be located at the boundary of the box constraint. 
In principle, the active set idea can be applied to accelerate any algorithm for solving problem \eqref{eq.prob1.multi} which does not exploit the special structure of its solution. Since the random CD algorithm is the state-or-the-art algorithm for solving problem \eqref{eq.prob1.multi}, this paper devotes to further accelerate the random CD algorithm by the active set strategy. Below, we first introduce the random CD algorithm in details.


\subsection{Random CD Algorithm}
Random permuted CD \cite{zhilin_icc2019} is one of the most efficient variants of the CD type of algorithms for solving problem \eqref{eq.prob1.multi}. 
At each iteration, the algorithm randomly permutes the indices of all variables and then updates the variables one by one according to the order in the permutation. In particular, for any given coordinate $(b,n),$ we need to solve the following one-dimensional problem 
\begin{subequations}\label{eq.prob4}
	\begin{alignat}{2}\label{eq.prob4.1}
		&\underset{d}{\operatorname{minimize}}    &\quad& \sum_{j=1}^B\Bigg(\log\left(1+dg_{jbn}\mathbf{s}_{bn}^H\boldsymbol\Sigma_j^{-1}\mathbf{s}_{bn}\right)\nonumber\\
		&&&\qquad\qquad\left.-\frac{dg_{jbn}\mathbf{s}_{bn}^H\boldsymbol\Sigma_j^{-1}\hat{\boldsymbol\Sigma}_j\boldsymbol\Sigma_j^{-1}\mathbf{s}_{bn}}{1+dg_{jbn}\mathbf{s}_{bn}^H\boldsymbol\Sigma_j^{-1}\mathbf{s}_{bn}} \right)\\
		&\operatorname{subject\,to} &      &d \in [-a_{bn},1-a_{bn}]
		\label{eq.prob4.3}
	\end{alignat}
\end{subequations} in order to possibly update $a_{bn}.$ The closed-form solution for the above problem does not exist, which is different from the single-cell case. However, as the derivative of the objective function of \eqref{eq.prob4} involves a polynomial function of degree $2B-1,$ we can find the desired $d$ by a root-finding algorithm. The CD algorithm is terminated when 
\begin{equation}\label{terminate}
\|\operatorname{Proj}(\underline{\textbf{a}} - \nabla f(\underline{\textbf{a}})) - \underline{\textbf{a}}\|_2 < \varepsilon
\end{equation} is satisfied, where $\varepsilon > 0$ is the error tolerance. The details of the random CD algorithm are summarized in Algorithm~1.

\begin{center}
	\framebox{
		\begin{minipage}{8.3cm}
			\textbf{Algorithm 1: Random CD algorithm for solving problem \eqref{eq.prob1.multi}}
			\liuhao{\begin{algorithmic}[1]
					\STATE \textbf{Initialize:} $\underline{\textbf{a}} = \mathbf{0},$ $\boldsymbol\Sigma_b^{-1} = \sigma_w^{-2}\textbf{I},$ $b = 1,\ldots B,$ and $\varepsilon > 0;$
					\REPEAT[\emph{one iteration}]
					\STATE Randomly select a permutation $\{ i_1, i_2, \ldots i_{BN} \}$ of the coordinate indices $\{1, 2, \ldots BN\}$ of $\underline{\textbf{a}};$
					\FOR{$n = 1$ to $BN$}
					\STATE Solve problem \eqref{eq.prob4} to obtain $d;$
					\STATE $a_{i_n} \leftarrow a_{i_n} + d;$
					\STATE $\boldsymbol\Sigma_b^{-1} \leftarrow \boldsymbol\Sigma_b^{-1} - d  \frac{\boldsymbol\Sigma_b^{-1} \textbf{s}_{i_n} \textbf{s}_{i_n}^H \boldsymbol\Sigma_b^{-1}}{1 + d \textbf{s}_{i_n}^H \boldsymbol\Sigma_b^{-1} \textbf{s}_{i_n}},$ $b = 1, \ldots ,B;$
					\ENDFOR
					\UNTIL{$\|\operatorname{Proj}(\underline{\textbf{a}} - \nabla f(\underline{\textbf{a}})) - \underline{\textbf{a}}\|_2 < \varepsilon$}
					\STATE \textbf{Output:}  $\underline{\textbf{a}}.$
			\end{algorithmic}}
			\label{random_cd}
		\end{minipage}
	}
\end{center}

\subsection{Proposed Algorithm}
Notice that, at each iteration, Algorithm 1 treats all coordinates equally and tries to update all of them. However, due to the special structure of the solution of problem \eqref{eq.prob1.multi}, there will be quite a lot of coordinates $(b,n)$ with which problem \eqref{eq.prob4} has been solved but $a_{bn}$ does not change, i.e., the corresponding solution of problem \eqref{eq.prob4} is zero. This kind of computation is unnecessary and slows down Algorithm 1. The goal of this paper is to use the active set selection strategy to reduce the unnecessary computations and accelerate Algorithm 1.
%



%

In contrast to Algorithm~1, at each iteration, the active set CD algorithm first judiciously selects an active set, then uses Algorithm~1 to update the coordinates in the active set \emph{once}.
Due to the special structure of the solution of problem \eqref{eq.prob1.multi}, it is expected that the cardinality of the selected active set will be significantly less than the total number of devices (i.e., $BN$), and more importantly will gradually decrease as the iteration goes on. Therefore, the active set CD algorithm will save much unnecessary computational cost in Algorithm~1 and significantly accelerates it.


\subsubsection{Active Set Selection} In principle, the desirable active set should contain coordinates that contribute the most to the deviation from the optimality condition in \eqref{eq.kkt} at each iteration, so that their updates would improve the objective function as much as possible; on the other hand, its cardinality should be as small as possible in order to avoid unnecessary computation and improve the computational efficiency.
Therefore, there is a trade-off in selecting the coordinates that violate \eqref{eq.kkt} into the active set. 
In practice, our selection strategy of the active set $\mathcal{A}^k$ at a given feasible point $\underline{\mathbf{a}}^k$ is mainly based on the degree of the violation of the first-order optimality condition \eqref{eq.kkt}.
Specifically, at the $k$-th iteration, (i) in case $a^k_{bn} = 0,$ we only choose coordinates with a sufficiently large negative gradient into the active set; (ii) similarly, in case $a^k_{bn} = 1,$ we choose coordinates with a sufficiently large positive gradient into the active set; and (iii) in case $a^k_{bn} \in(0,1) ,$ we choose coordinates with the absolute value of the gradient being large enough into the active set.
Mathematically, the proposed selection strategy of the active set $\mathcal{A}^k$ can be expressed as
\begin{align}\label{selectionrule}
	\mathcal{A}^k = & \left\{ (b,n) \mid a^k_{bn} = 0 \ \textup{and} \ \nabla f^k_{bn} < -\delta_{bn}^k \right\}\nonumber \\
	& \cup \left\{ (b,n) \mid a^k_{bn} = 1 \ \textup{and} \ \nabla f^k_{bn} > \delta_{bn}^k \right\} \\
	& \cup \left\{ (b,n)\mid 0 < a^k_{bn} < 1 \ \textup{and} \ \left|\nabla f^k_{bn}\right| > \delta_{bn}^k \right\},\nonumber
\end{align}
where $\nabla f_{bn}^k$ denotes $\left[\nabla f(\underline{\mathbf{a}}^k)\right]_{bn}$ and $\bm{\delta}^k \in \mathbb{R}^{BN}_+$ is a threshold vector which changes with interation.
 
The choice of the threshold vector $\bm{\delta}^k$ in \eqref{selectionrule} plays an important role in balancing the competing goals of decreasing the objective function and reducing the cardinality of the active set (thus the computational cost) at the $k$-th iteration. If the components of $\bm{\delta}^k$ are large, the cardinality of the active set at the $k$-th iteration will be small (and the iteration will need less computations), but the decrease of the objective function will also be small.
The reverse is also true. In particular, if $\bm{\delta}^k=\textbf{0},$ all coordinates that violate \eqref{eq.kkt} will be chosen in the active set. 
A way of choosing the parameter $\bm{\delta}^k$ is to set a relatively large initial $\bm{\delta}^0$ and then gradually decrease $\bm{\delta}^k$ at each iteration. 




\subsubsection{Active Set CD Algorithm}
Based on the above analysis, we propose the active set CD algorithm for solving problem \eqref{eq.prob1.multi}, which is detailed in Algorithm~2.

\begin{center}
	\framebox{
		\begin{minipage}{8.3cm}
			\textbf{Algorithm 2: Proposed active set CD algorithm for solving problem \eqref{eq.prob1.multi}}
			\liuhao{\begin{algorithmic}[1]
					\STATE \textbf{Initialize:} $\underline{\textbf{a}}^0 = \mathbf{0},$ $k = 0,$ and $\varepsilon > 0;$ 
					\REPEAT[\emph{one iteration}]
					\STATE Update $ \bm{\delta}^k ;$
					\STATE Select the active set $\mathcal{A}^k$ according to \eqref{selectionrule}; 
					\STATE Apply Algorithm~1 to update all coordinates in $\mathcal{A}^k$ \emph{only once;}
					\STATE 
					Set $k \leftarrow k+1;$
					\UNTIL{$\|\operatorname{Proj}(\underline{\textbf{a}}^k - \nabla f^k) - \underline{\textbf{a}}^k\|_2 < \varepsilon$}
					\STATE \textbf{Output:}  $\underline{\textbf{a}}^k.$
			\end{algorithmic}}
			\label{alg_act}
		\end{minipage}
	}
\end{center}

The convergence property of the proposed active set CD Algorithm~2 is presented in the following Theorem \ref{theorem1}. Note that a poor choice of the active set might lead to oscillation and even divergence of the corresponding algorithm. The following finite termination property of the proposed Algorithm~2 is mainly because of the active set selection strategy in \eqref{selectionrule} (and careful choices of vectors $\bm{\delta}^k$), and the convergence property of Algorithm 1.
Due to the space limitation, we omit the rigorous proof of Theorem~\ref{theorem1}.
\begin{theorem} \label{theorem1}
	For any given error tolerance $\varepsilon>0,$ suppose that the threshold vector $\bm{\delta}^k$ in \eqref{selectionrule} satisfy that $\| \bm{\delta}^k \|_2$ decreases monotonically with $k$ and $\lim\limits_{k \rightarrow \infty} \| \bm{\delta}^k \|_2 < \varepsilon ,$ then the proposed active set CD Algorithm~2 will terminate within a finite number of iterations.
\end{theorem}

Finally, it is worth mentioning that the motivation of Algorithm~2 in this paper is similar to that of \cite{wang2021efficient}, which also uses the sporadic nature of the device activities to lower the complexity in algorithmic design.
However, the philosophy of selecting the active set in the two algorithms substantially differs from each other. In particular, the active set in \cite{wang2021efficient} aims to contain all the active devices, while the active set in Algorithm~2 aims to contain devices whose activities are most likely to be incorrect. The reason for the difference is that the large-scale fading coefficients are assumed to be known in this paper, so that $\textbf{0} \leq \underline{\textbf{a}} \leq \textbf{1},$ but in \cite{wang2021efficient} the large-scaling fading coefficients need to be estimated. This difference further leads to a totally different convergence behavior of the active set in the two algorithms: the cardinality of the active set in \cite{wang2021efficient} will converge to a constant (which is in the same order as the number of active devices) but the cardinality of the active set in Algorithm~2 will converge to zero, i.e., there will be no (uncertain) devices which will violate the optimality condition when the algorithm terminates.

%
%

%
%

%
%

%
%

%

%
%
%
%
%
%

%
%
%
%
%
%
%
%
%
%
%
%
%
%
%
%
%
%

%
%
\section{Simulation Results}
In this section, we demonstrate the efficiency of our proposed active set CD algorithm via numerical simulations. The parameters in simulations are the same as those in \cite{chen2021sparse}.
More specifically, we consider a multi-cell system consisting of $B = 7$ hexagonal cells with wrap-around at the boundary, and $N$ devices uniformly distributed within each cell.
The radius of each cell is $250$m;
the channel path-loss is modeled as $128.1+37.6\log_{10}(d),$ where $d$ is the corresponding BS-device distance in km; the transmit power of each device is set as $23$dBm, and the background noise power is $-169$dBm/Hz over $10$MHz; 
and the number of antennas at each BS is $M = 512.$
In all simulations, all signature sequences are sampled from i.i.d.\ complex Gaussian distribution with zero mean and unit variance.
The tolerance is $\varepsilon = 10^{-3}$ and in the proposed Algorithm 2, the threshold vector $\bm{\delta}^k$ is chosen as
\begin{align*}
	\delta_{bn}^k = 
	\begin{cases}
		10^{-k-2} \times \max \left\{ \left| \nabla f^k_{bn} \right| \mid a_{bn}^k = 0 \right\}, & \operatorname{if}~a_{bn}^k = 0;   \\
		10^{-k-2} \times \max \left\{ \left| \nabla f^k_{bn} \right| \mid a_{bn}^k = 1 \right\}, & \operatorname{if}~a_{bn}^k = 1; \\
 		\max\{ 5^{-k}, \varepsilon/\sqrt{0.3BN} \}, & \operatorname{otherwise,}
	\end{cases}
\end{align*}
which satisfies the conditions in Theorem \ref{theorem1}. 

\begin{figure}[t]
	\centering
	\includegraphics[scale = 0.55]{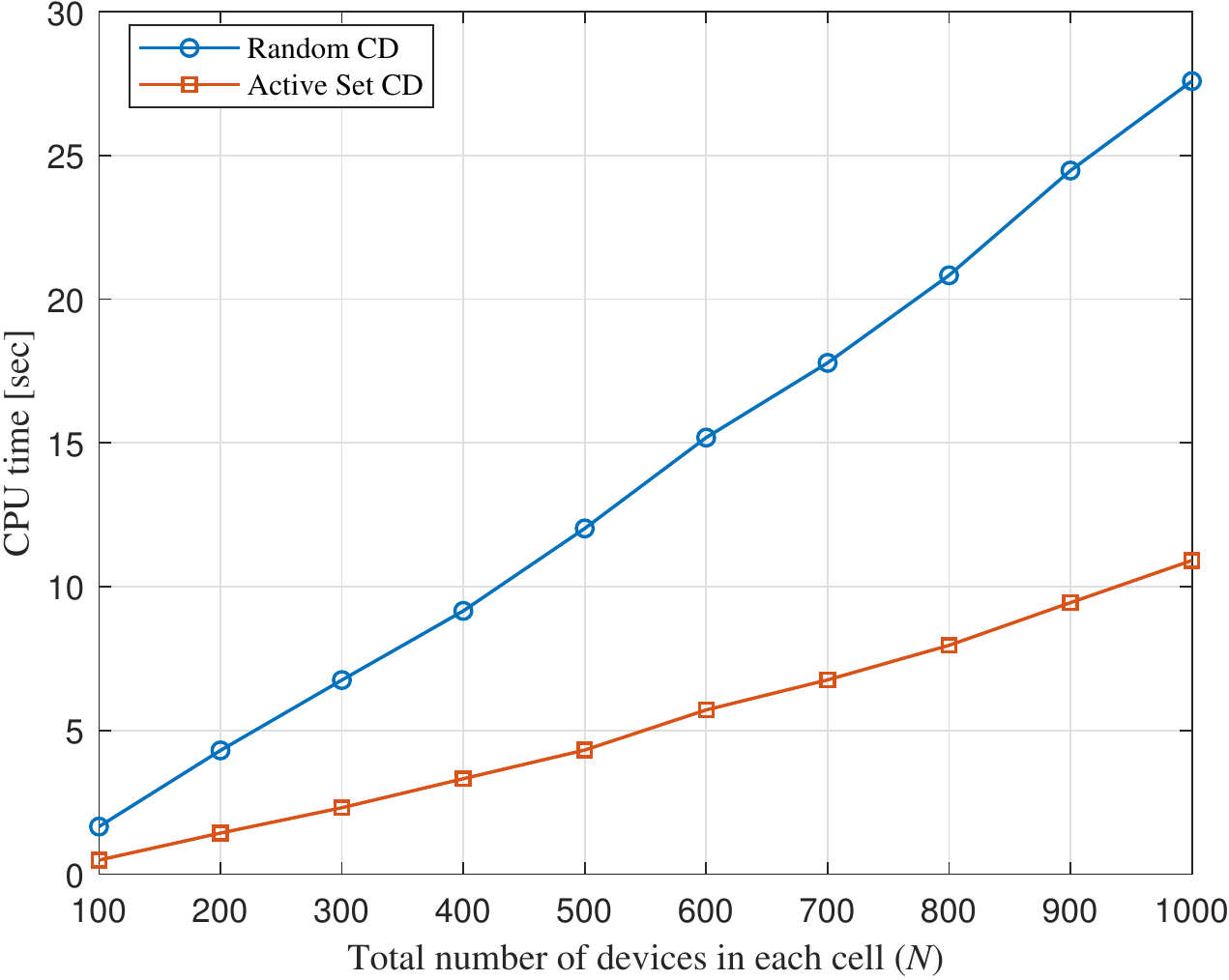}
	\caption{Average CPU time comparison of the proposed active set CD algorithm and the random CD algorithm.}
	\label{cputime}
\end{figure}

\begin{figure}[t]
	\centering
	\includegraphics[scale = 0.55]{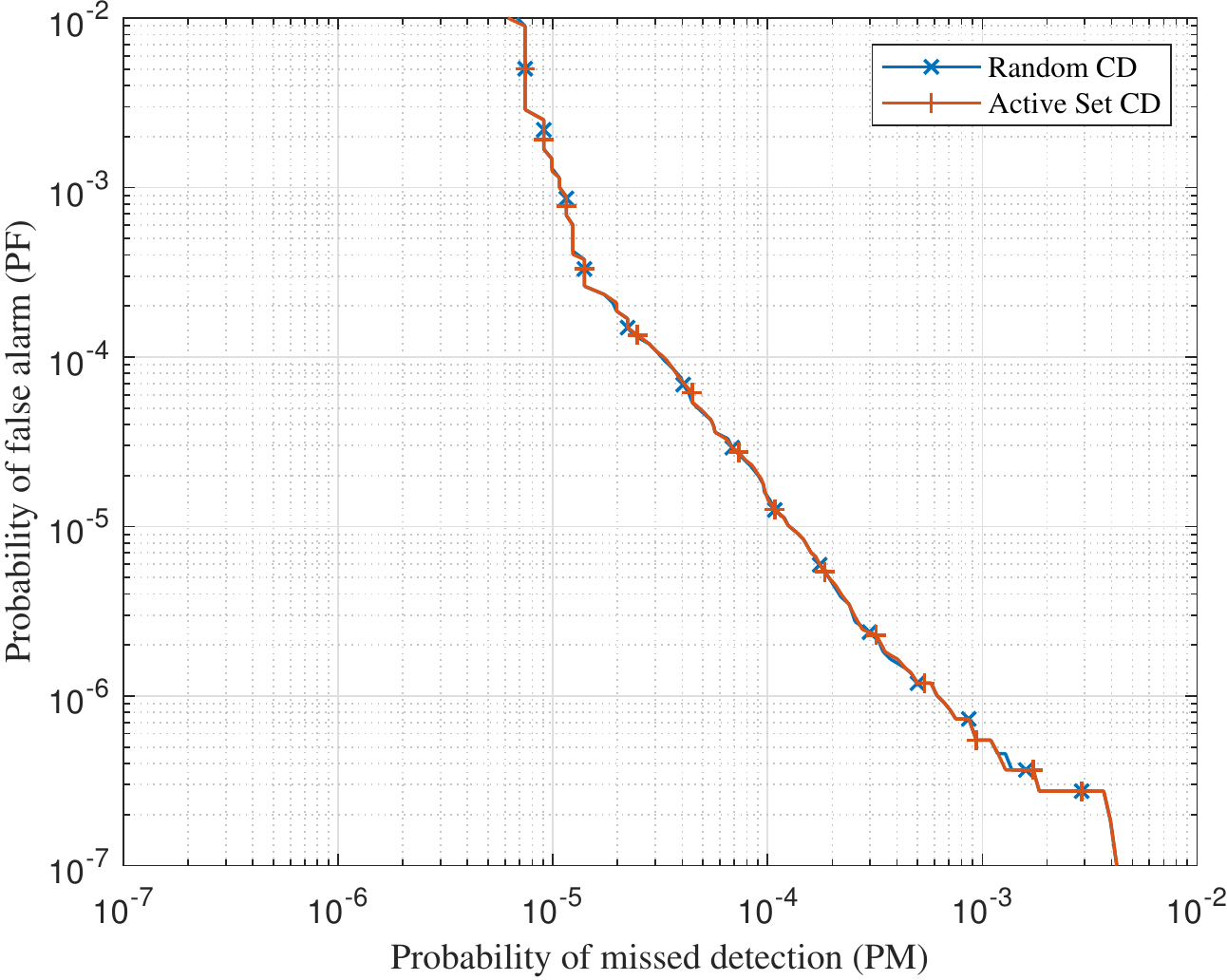}
	\caption{The detection performance of the proposed active set CD algorithm and the random CD algorithm.}
	\label{detection}
\end{figure}

Fig.~\ref{cputime} plots the CPU time comparison of Algorithm~1 and Algorithm~2 versus the number of devices in each cell.
In each cell, for a given number of devices $N,$ we set the number of active devices $K = N/10$ with $L$ ranging from $15$ (for $N=100$) to $45$ (for $N=1000$). The simulation results are obtained by averaging over $200$ Monte-Carlo runs. It can be observed that the active set CD algorithm is significantly more efficient than the random CD algorithm.

We now compare the detection performance (i.e., probability of missed detection and probability of false alarm) of the two algorithms in Fig.~\ref{detection} with $N = 200,$ $K = 20,$ and $L = 20.$
A trade-off between missed detection and false alarm can be obtained by selecting different thresholds.
We can observe from Fig.~\ref{detection} that the detection performance curves of these two algorithms completely overlap, which shows that both the active set CD algorithm and the random CD algorithm converge to the same solution of problem \eqref{eq.prob1.multi} (albeit the problem is generally nonconvex).

\begin{figure}[t]
	\centering
	\subfigure[]{
		\includegraphics[height=0.14\textheight]{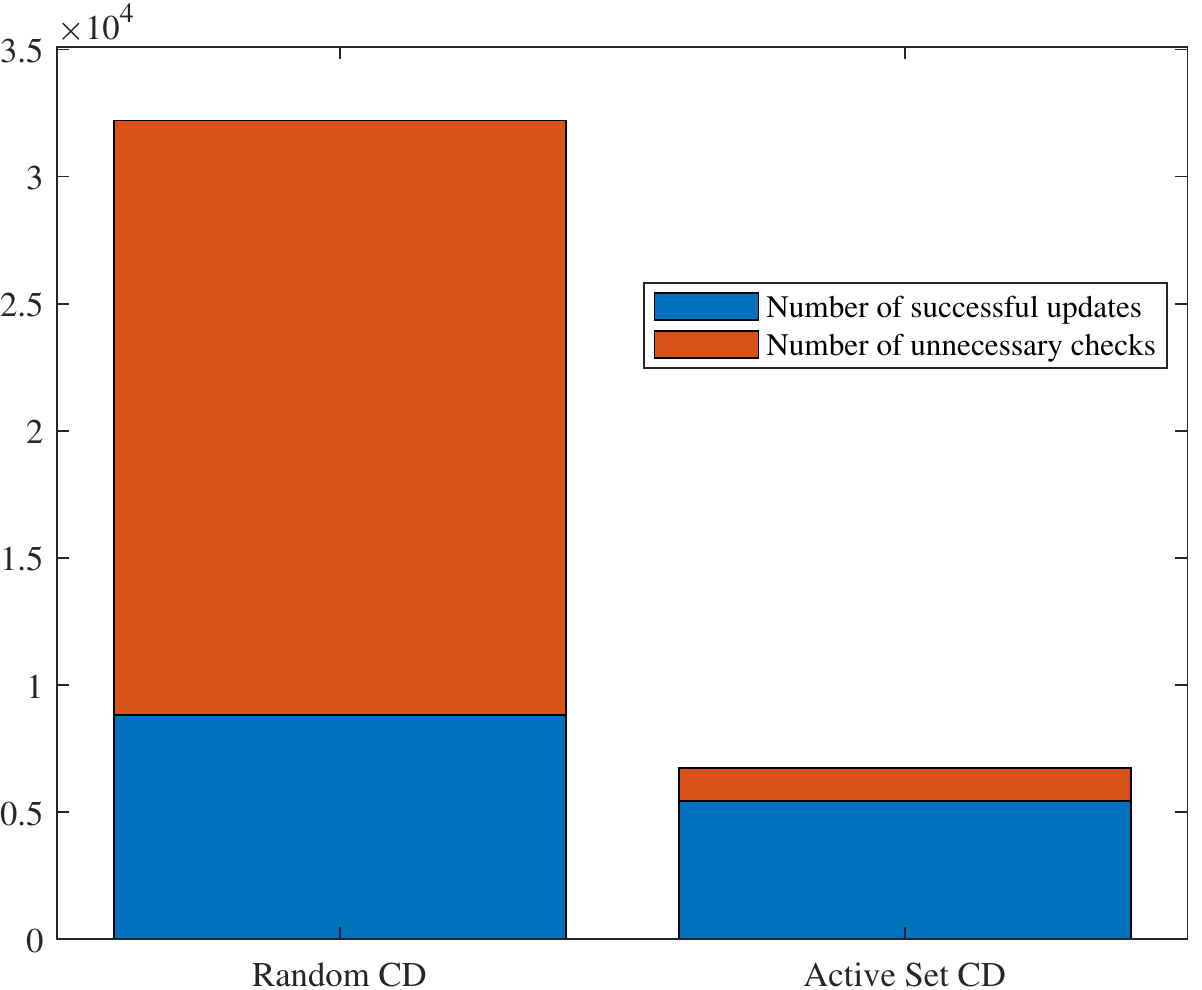}
		\label{number:update}
	}
	\subfigure[]{
		\raisebox{-0.03\height}{\includegraphics[height=0.14\textheight]{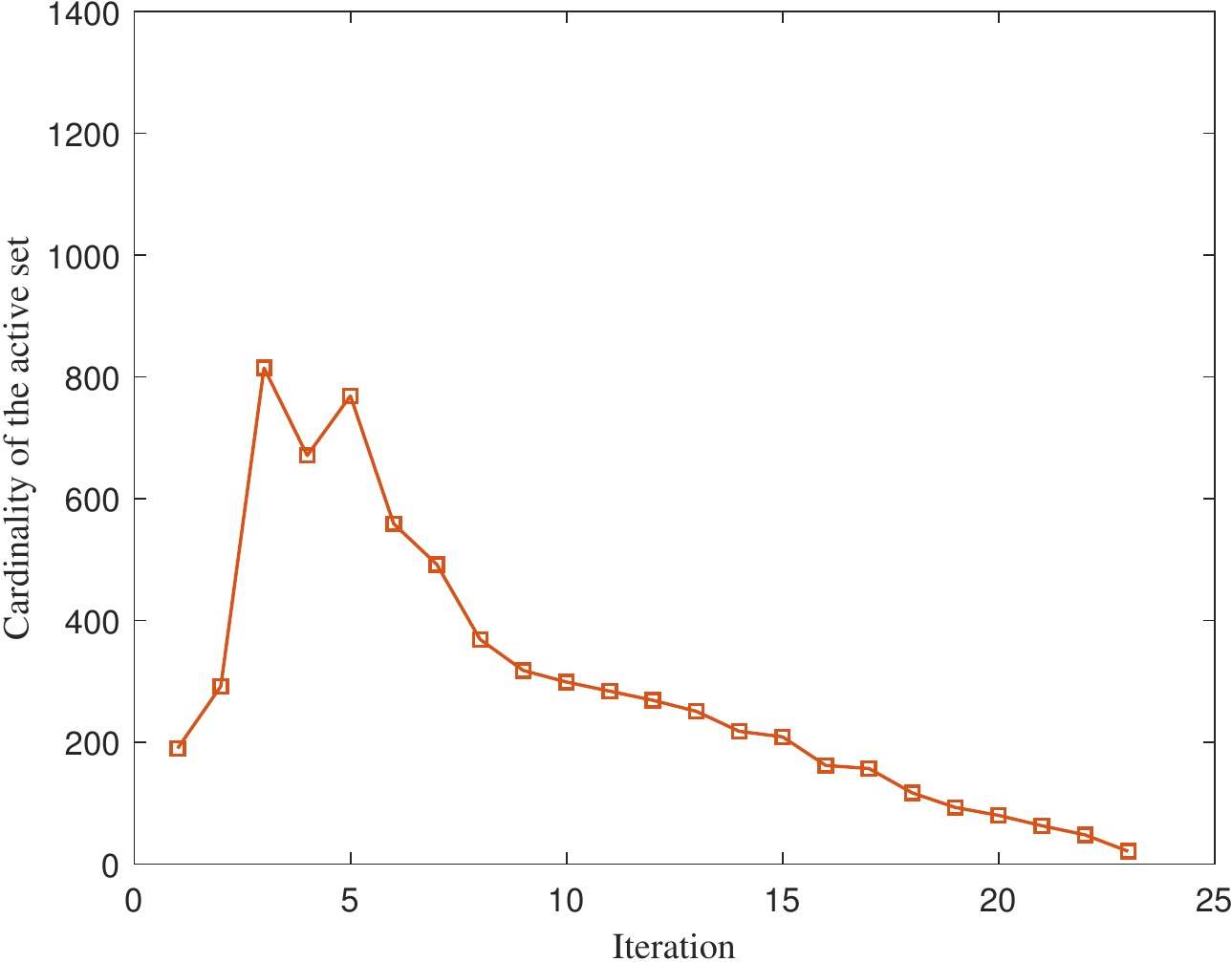}}
		\label{number:activeset}
	}
	
	\caption{Behaviors of the proposed active set algorithm: (a) The number of times of successful updates and unnecessary checks; (b) The cardinality of the active set over the iteration.}
\end{figure}


To reveal why and how the active set selection helps significantly reduce the computational cost, we present more details on the update of each variable and classify it into two different cases: case (i) where the variable is successfully updated and case (ii) where unnecessary check/computation is done (and the variable is not updated). The dominant computational cost in case (i) consists of solving the roots of the polynomial function with degree $2B-1$ and doing rank-one updates to the matrix $\boldsymbol\Sigma_b^{-1}$ for all $b$ and the dominant computational cost in case (ii) is to solve the roots of the polynomial function. 

Fig.~\ref{number:update} plots the numbers of cases (i) and (ii) that happened in different algorithms on a (randomly generated) problem instance with $N = 200,$ $K = 20,$ and $L = 20.$ It can be clearly seen from Fig.~\ref{number:update} that the number of unnecessary checks is significantly greater than the number of successful updates in the random CD algorithm, which shows that a lot of unnecessary computations are done in it. This is not surprising, as many entries of the solution of problem \eqref{eq.prob1.multi} are either $0$ or $1$ but the random CD algorithm fails to exploit this special structure by treating all entries equally. We can also observe from Fig.~\ref{number:update} that the proposed active set CD algorithm can significantly reduce the number of unnecessary checks thus significantly improving the computational efficiency of the random CD algorithm. This shows the importance of exploiting the special structure of problem \eqref{eq.prob1.multi} as well as the efficiency of the proposed active set selection strategy.


Fig.~\ref{number:activeset} shows the cardinality of the active set at each iteration in the proposed active set CD algorithm. We can observe from Fig.~\ref{number:activeset} that the cardinality of the active set gradually decreases and becomes zero when the algorithm terminates, which is in sharp contrast to that in \cite{wang2021efficient}.
Fig.~\ref{number:activeset} also demonstrates the active set CD algorithm can converge quickly 
and verifies the finite termination property of the algorithm in Theorem~\ref{theorem1}.



%

\section{Conclusions}

This paper proposes an efficient active set CD algorithm for solving covariance based device activity detection in a cooperative multi-cell massive MIMO system. The proposed algorithm selects those coordinates that can potentially most improve the objective function in the active set and only updates the coordinates in the active set at each iteration, which is in contrast to the random CD algorithm which updates all coordinates at each iteration. Numerical results show that the proposed active set CD algorithm can achieve the same detection performance as the state-of-the-art random CD algorithm but with a significant less CPU time.

\ifCLASSOPTIONcaptionsoff
	\newpage
\fi

\bibliographystyle{IEEEtran}
\bibliography{IEEEabrv,chenbib20210424}


%







\end{document}